      [1999/12/01 v1.4c Il Nuovo Cimento]
\documentclass{cimento}
\usepackage{atlasphysics}
\title{ Top-Antitop cross section measurement in the di-lepton decay channel with ATLAS }
\author{D.B.~Ta\from{ins:x}\ETC}
\instlist{\inst{ins:x} Physikalisches Institut, University of Bonn, Germany }
\PACSes{\PACSit{14.65Ha}{Top Quarks}}
\begin{document}

\maketitle

\begin{abstract}
We present simulations of the production cross-section measurement of
top-antitop pairs in the di-leptonic decay channel with the ATLAS detector. 
This study uses the Commissioning Service Challenge (CSC) data, which is the 
latest and centrally produced Monte-Carlo data set to validate the detector 
simulation before the actual data taking. The signal process was generated 
with MC@NLO~\cite{Frixione:2006gn} and important background processes 
were studied. A cut and count method and 
two likelihood methods were employed to measure the cross section and important 
systematic effects
were investigated. The expected statistical and systematic errors for a 
luminosity of 100~$\ipb$ are also given.
\end{abstract}



\section{Di-lepton signature and background processes}
The characteristic di-lepton signature is given by two opposite charged 
leptons\footnote{The lepton identification is decribed in Ref.~\cite{ATLASdpaper}. Additionally isolation from
energy depositions in a hollow cone of $\Delta R  <  0.2$ of 6~\GeV ~is required for electrons and
isolation from reconstructed jets within $\Delta R  <  0.2$ is required for muons.}
from the \Wboson decay and at least two jets from the decay of the two b-quarks as well as a large amount of
missing transverse energy (\met) from the neutrinos.



The leptonically decaying \Zboson in Drell-Yan events is an important background process for the di-leptonic
subchannels with same flavor leptons. 
A very similar signature to the signal is produced by di-boson ($W/Z$) decays. 
Background processes that arise by misidentified leptons are mainly \ttbar lepton+jets, $W$+jets and QCD background events.
Also single top events have been investigated.



%
\section{Cross section measurement methods}
\subsection{Cut and count method}
A simple cut based analysis on the lepton and jet \pt ~and on \met ~was optimized for the best $S/\sqrt{S+B}$. 
Events with same flavored leptons and an invariant di-lepton
mass compatible with the \Zboson mass were rejected. The optimized results require cuts of \pt$>$20~\GeV ~for all visible objects 
and \met$>$35~\GeV ~for $ee$ and $\mu\mu$ channel and 25~\GeV ~for the $e\mu$ channel.


\subsection{Likelihood method}
A second method employs the different shapes of signal and background in selected multidimensional distributions.
A likelihood fit can determine the fraction of signal and background 
events in the total sample by fitting to the total event shape.
The likelihood fit was performed with the distributions of 
the variables $|\Delta \varphi|$~between
the highest \pt ~lepton and the
\met ~vector, and $|\Delta \varphi|$~between
the highest \pt ~jet and the \met ~vector.

\subsection{Inclusive template method}
The inclusive template method uses templates based on the two-dimensional distributions of the \met
~and the jet multiplicity for the three dominant sources of $e\mu$ di-leptons, i.e. \ttbar
di-lepton, $\Zzero\ra\tau\tau$ and $WW$. This method requires leptons with tighter isolation
criteria and rejects events with \met ~energy aligned with the muon in order to minimize events with
misidentified leptons.
The final fit has ten variables including the cross sections for the three processes.

\section{Systematic uncertainties}
The effect of the jet energy scale uncertainty was estimated 
by rescaling all reconstructed jet vectors by $\pm 5\%$ and changing the missing transverse energy 
to preserve the total transverse momentum in the event. Initial and final state radiation was
investigated with fast simulated samples in which PYTHIA~\cite{Sjostrand:2007gs} ISR and FSR parameters were increased
by $100\%$ or halved. Finally the uncertainties from PDF variations were investigated by reweighting the events with 
the probability of both initial partons evaluated at the same $x_1$, $x_2$ and $Q^2$ value as in the generated event but with
the error PDFs provided by the CTEQ and MRST collaboration. The error was evaluated by the Hessian approach~\cite{Tung:2002vr}.

\section{Summary}

The following table summarizes the anticipated statistical and systematic errors evaluated at an integrated luminosity of $100~\ipb$. The full result will be published in the ATLAS CSC note.
\begin{eqnarray}
\mbox{Cut and Count method:} &\Delta \sigma /\sigma = & (4 (\mbox{stat}) {}^{+5}_{-2} 
(\mbox{syst}) \pm 2 (\mbox{pdf}) \pm 5 (\mbox{lumi})) \% \\ 
\mbox{Template method:} &\Delta \sigma /\sigma = & (4 (\mbox{stat}) \pm 4 (\mbox{syst}) \pm 2 (\mbox{pdf}) \pm 5 (\mbox{lumi})) \% \\ 
\mbox{Likelihood method method:} &\Delta \sigma /\sigma = & (5 (\mbox{stat}) {}^{+8}_{-5} (\mbox{syst}) 
\pm 2.4 (\mbox{pdf}) \pm 5 (\mbox{lumi})) \% 
\end{eqnarray}

\acknowledgments
The work on the inclusive template method was done by J. Sj\"olin, University of Stockholm. He also kindly provided the fast simulation samples for the ISR/FSR systematic studies. This work was supported by the Bundesministerium f\"ur Bildung und Forschung in the framework of FSP 101.


\begin{thebibliography}{0}
%
%
%
\bibitem{Frixione:2006gn}
  S.~Frixione and B.~R.~Webber,
    ``The MC@NLO 3.3 event generator,''
	   arXiv:hep-ph/0612272.


\bibitem{ATLASdpaper}
The ATLAS collaboration, 
``The ATLAS Experiment at the CERN Large Hadron Collider,''
2007/2008 submitted to JINST

\bibitem{Sjostrand:2007gs}
  T.~Sjostrand, S.~Mrenna and P.~Skands,
    ``A Brief Introduction to PYTHIA 8.1,''
	   Comput.\ Phys.\ Commun.\  {\bf 178}, 852 (2008)
		  [arXiv:0710.3820 [hep-ph]].


\bibitem{Tung:2002vr}
  W.~K.~Tung,
    ``New generation of parton distributions with uncertainties from global  QCD
	   analysis,''
		  Acta Phys.\ Polon.\  B {\bf 33} (2002) 2933
		    [arXiv:hep-ph/0206114].



\end{thebibliography}
\end{document}